\documentclass[a4paper,11pt]{article}
\usepackage{pos}

\usepackage{mathtools}
\usepackage{natbib}
\usepackage{fontenc}
\usepackage{times}
\usepackage[utf8]{inputenc}
\usepackage[polish, english]{babel}
\usepackage{polski}

\usepackage{lipsum}

\usepackage{plain}
\usepackage{natbib}

\usepackage{graphicx}
\usepackage{float}
\usepackage{caption}
\usepackage{subcaption}

\title{Analysis of capability of detection of
extensive air showers by simple scintillator
detectors}
 \ShortTitle{Results prediction for simple EAS detection}

\author*[1]{Jerzy Pryga}
\author[2]{Weronika Stanek}

\affiliation[1]{Jagiellonian University,\\
  ul. Gołębia 24, 31-007, Cracow, Poland}

\affiliation[2]{AGH University of Science and Technology,\\
al. Mickiewicza 30
30-059, Cracow, Poland}

\forColl{CREDO} % W/O "Collaboration"

\emailAdd{jerzy.pryga@student.uj.edu.pl}
\emailAdd{wstanek@student.agh.edu.pl}

\abstract{One of the main objectives of the CREDO project is to search for so-called Cosmic-Ray Ensembles (CRE) \cite{homola2020cosmic}. To confirm the existence of such phenomena a massive scale observation of even relatively low energy Extensive Air Showers (EAS) and an analysis of their correlations in time must be performed.  To make such observations possible, an infrastructure of widely spread detectors connected in a global network should be developed using low-cost devices capable of collecting data for a long period of time.
For each of these detectors or small detector systems the probability of detection of an EAS has to be determined. Such information is crucial in the analysis and interpretation of collected data.
In the case of large number of systems with different properties the standard approach based on detailed and extensive simulations is not possible, thus a faster method is developed. Knowing the characteristics of EAS from more general simulations any required probability is calculated using appropriate parameterization taking into account EAS spectrum, energy dependence of particle density and zenith angle dependence. This allows to estimate expected number of EAS events measured by a set of small detectors \cite{Karbowiak_2020} and compare results of calculations with these measurements. }

\FullConference{37$^{\rm{th}}$ International Cosmic Ray Conference (ICRC 2021)\\
		July 12th -- 23rd, 2021\\
		Online -- Berlin, Germany}

\begin{document}
\maketitle

\section{Introduction}

The main motivation for this work is to provide helpful information about one of methods of detection of Extensive Air Showers (EAS) proposed for Cosmic Rays Extremely Distributed Observatory (CREDO) \cite{homola2020cosmic}. As the goal of this project, which is a search of cosmic rays ensembles, require monitoring of cosmic rays and EAS on a global scale, demand for a simple, cheap and reliable detector rises. One of proposed ideas is analysed in this work: several scintillator detectors connected in a coincidence system. When several devices give signal at almost the same time it indicates an occurrence of a cosmic ray cascade. Performed analysis aims to give answer if such very simple systems can be reliable and efficient enough to use in a bigger network of devices and collect useful data.

\section{Analysis}

\subsection{Simulations}
\label{simulations}

First step in the analysis was to perform shower simulations using CORSIKA software \cite{heck1998corsika}. These data is later analysed to characterise density of particles form the shower on the ground as a function of cascade parameters. 
Currently simulations of protons as a primary particle are used, with 18 different energies starting from 1 TeV up to 4000 TeV. For now, no heavier nuclei were included but it is planned. For each energy 1000 to 20 000 cascades were simulated. To examine angular distribution of produced particles, simulations of cascades at several selected zenith angles with energy of 100 TeV were analysed.

\subsection{Analysed system}
\label{analysed system}

Systems which are analysed in this work consists of several flat scintillator detectors connected in a coincidence. It is assumed that particles can be detected from all directions and different types of them can not be distinguished. Time interval in which two or more signals from different devices are considered a coincidence is called here \textit{coincidence time} $\delta T$. Other properties of considered detectors are their \textit{area of the surface} $A$, \textit{efficiency} $\eta$ and \textit{frequency of non cosmic background signals} $f_{bg}$. All devices in the system are identical and are placed very close to each other. With those properties it is possible to analyse such systems in easy and relatively realistic way.

\subsection{Background}
\label{background}

Starting point in the analysis is to evaluate flux of background particles, and probability of fake signals i.e. not caused by EAS. In this work, term \textit{background} means uncorrelated, single cosmic-rays which comes from all direction with constant flux indicated as $I_{bg}$.

For short measurement time $\delta T$ the probability of a fake signal to happen is approximately:

\begin{equation}
    P_{bg} =  1 - \exp\left(-\delta T \left(\eta \cdot A \cdot I_{bg}+f_{bg} \right)\right)
    \label{eq:P_bg}
\end{equation}

After that, chance of $k$ coincidence signals is computed for system made of $n$ detectors. With probability of detection for each device equal $P_{bg}$ it is denoted as $Q\left(n, k, P_{bg}\right)$ and given by binomial distribution. Expected number of such events in the \textit{time of measurement} $T$ is computed using following formula:

\begin{equation}
    \langle N_{bg}\left(k\right) \rangle = Q\left(n, k, P_{bg}\right) \cdot \frac{T}{\delta T}
    \label{eq:N_bg}
\end{equation}

\subsection{Signals from the cascades}
\label{signals from the cascades}

Before starting any fitting or calculations some assumptions about EAS has to be taken. All showers are treated as circularly symmetrical and produced by protons. Because protons constitute about 78 \% of primary cosmic rays, the  spectrum for all cosmic-ray particles was used. Frequency of those protons is constant in time and indicated here as $j\left(E\right)$ \cite{aartsen2013measurement}. This function is obviously a model confirmed by various experiments \cite{particle1998review}.

The next step is to extract information from simulations. The easiest way to characterise EAS is by is particles density $\rho$. In this work it is a function of parameters such as energy of primary particle $E$, distance from its centre $r$, zenith angle of primary particle approach direction $\theta$ and total number of produced particles $N_{part}$. Each of these quantities were analysed more or less exhaustively and pre-defined functions were fitted to data.

Final $\rho$ function is a product of all this factors given by following formula:

\begin{equation}
    \rho\left(r, \theta, E, N_{part}\right) = \rho_{norm}\left(r\right) \cdot F_{\theta}\left(\theta\right) \cdot F_E\left(E, r\right) \cdot F_N\left(N_{part}, r\right)
    \label{eq:rho}
\end{equation}

\begin{itemize}
    \item $\rho_{norm}\left(r\right)$ - The most important factor which represents relation between particle density and distance from the centre of the shower. This function is fitted for certain vertical cascade of chosen energy to which other factors are normalised. Figure \ref{fig:rho_r} presents shape of this function.
    \begin{figure}[h!]
        \centering
        \begin{subfigure}[b]{0.47 \textwidth}
            \centering
            \includegraphics[width=\textwidth]{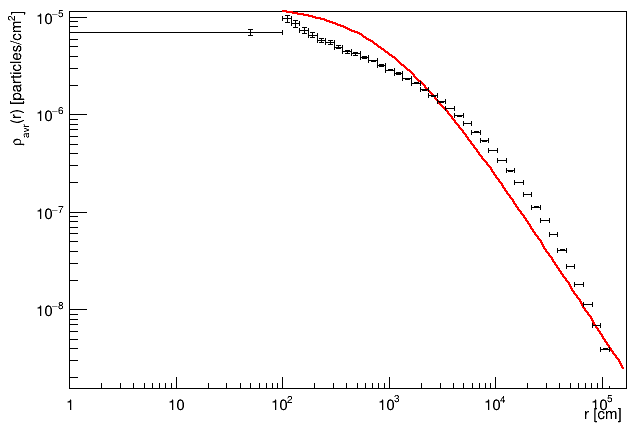}
            \caption{}
            \label{fig:rho_r_mu}
        \end{subfigure}
        \begin{subfigure}[b]{0.47 \textwidth}
            \centering
            \includegraphics[width=\textwidth]{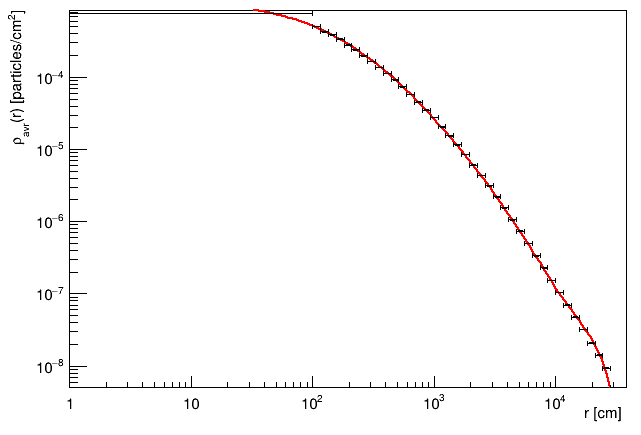}
            \caption{}
            \label{fig:rho_r_gamma}
        \end{subfigure}

        \caption{$\rho \left( r \right)$ function fitted to data for \ref{fig:rho_r_mu} muons and \ref{fig:rho_r_gamma} photons.}
        \label{fig:rho_r}
    \end{figure}
    
    \item $F_{E}\left(E, r\right)$ - This factor scales the density with different energy and modifies its distance distribution. It is presented on figure \ref{fig:rho_rE}.
    \begin{figure}[h!]
        \centering
        \begin{subfigure}[b]{0.47 \textwidth}
            \centering
            \includegraphics[width=\textwidth]{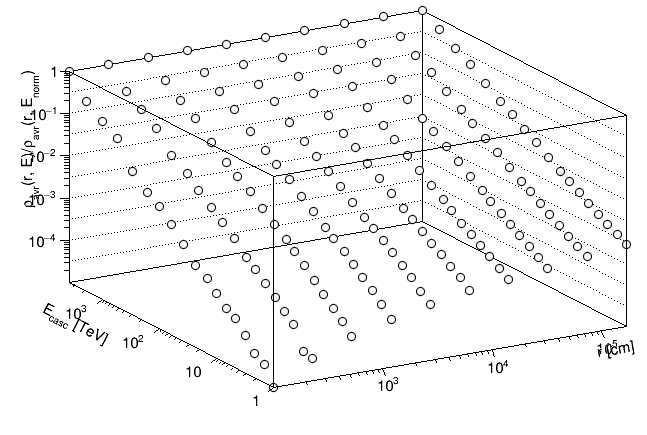}
            \caption{}
            \label{fig:rho_rE_mu}
        \end{subfigure}
        \begin{subfigure}[b]{0.47 \textwidth}
            \centering
            \includegraphics[width=\textwidth]{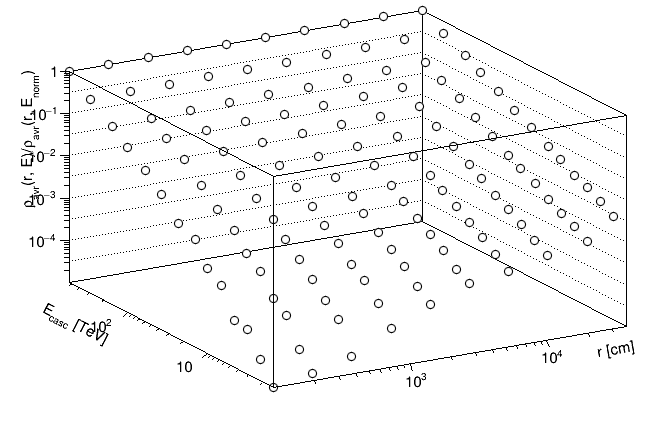}
            \caption{}
            \label{fig:rho_rE_gamma}
        \end{subfigure}

        \caption{Graph of $F_{E}\left(E, r\right)$ for \ref{fig:rho_rE_mu} muons and \ref{fig:rho_rE_gamma} photons.}
        \label{fig:rho_rE}
    \end{figure}
    
    \item $F_{N}\left(N_{part}, r\right)$ - Scales the density and its distance distribution as total number of produced particles changes. Their amount is strongly connected with the altitude at which the cascade started to form, but $N_{part}$ is a more convenient parameter to work with. It must be mentioned that their form vary for different particle types and currently for electromagnetic component of the showers $F_{N}$ factor is just a linear proportion of $\frac{N}{N_{avr}}$ without dependence on distance $r$. It is presented in Figure \ref{fig:rho_rN}.
    \begin{figure}[h!]
        \centering
        \begin{subfigure}[b]{0.47 \textwidth}
            \centering
            \includegraphics[width=\textwidth]{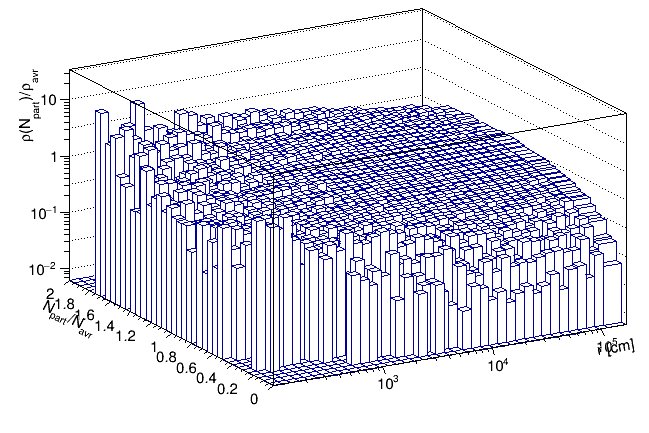}
            \caption{}
            \label{fig:rho_rN_mu}
        \end{subfigure}
        \begin{subfigure}[b]{0.47 \textwidth}
            \centering
            \includegraphics[width=\textwidth]{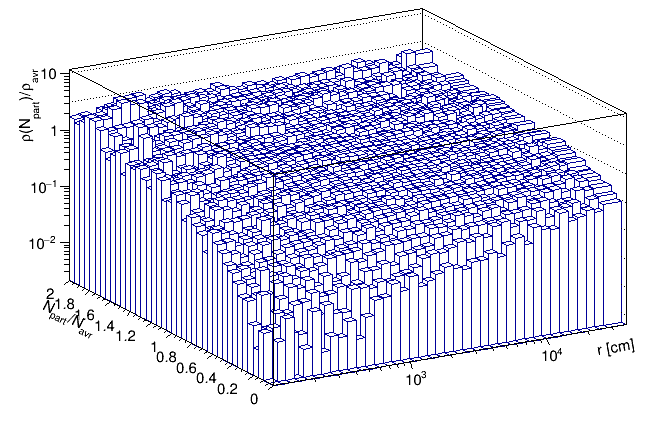}
            \caption{}
            \label{fig:rho_rN_gamma}
        \end{subfigure}

        \caption{Histograms of $F_{N}\left(N_{part}, r\right)$ for \ref{fig:rho_rN_mu} muons and \ref{fig:rho_rN_gamma} photons.}
        \label{fig:rho_rN}
    \end{figure}
    
    \item $F_{\theta}\left(\theta\right)$ - This factor scales the density as zenith angle of primary particle approach direction changes. Here a simplification was made. Namely, the shape of shower footprint is not circularly symmetrical as this angle increase. However, EAS comes from all directions with equal probability thus this effect can be neglected. Figure \ref{fig:rho_theta} presents its shape.
    \begin{figure}[h!]
        \centering
        \begin{subfigure}[b]{0.47 \textwidth}
            \centering
            \includegraphics[width=\textwidth]{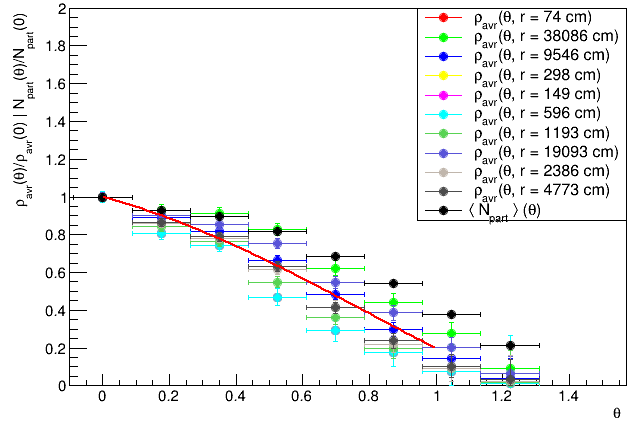}
            \caption{}
            \label{fig:rho_theta_mu}
        \end{subfigure}
        \begin{subfigure}[b]{0.47 \textwidth}
            \centering
            \includegraphics[width=\textwidth]{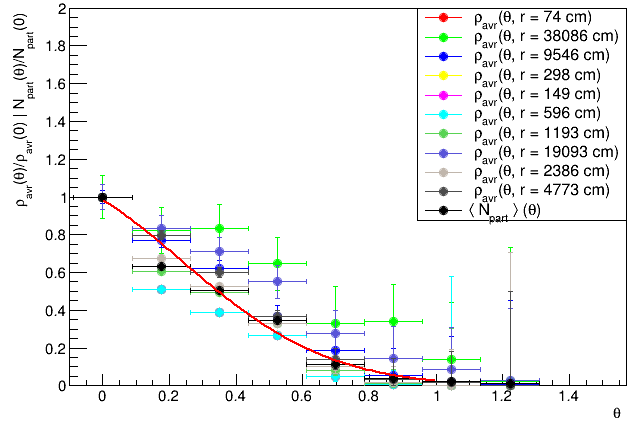}
            \caption{}
            \label{fig:rho_theta_gamma}
        \end{subfigure}

        \caption{$F_{\theta}\left(\theta\right)$ function fitted to data for \ref{fig:rho_theta_mu} muons and \ref{fig:rho_theta_gamma} photons.}
        \label{fig:rho_theta}
    \end{figure}
    
\end{itemize}

Next step is to calculate probability of a signal to happen when there is a cascade near the detector. Formula for a single signal probability is as follows:

\begin{equation}
    P = 1 - \exp\left(-\eta \cdot A \cdot \rho\left(r, \theta, E, N_{part}\right)\right)
    \label{eq:P}
\end{equation}

There is also another effect in the cascades that could have impact on the probability of coincidence signals. It is currently under investigation, but early results suggest that when one particle that originates from the shower is detected, then probability of finding another one in close region of such event should increase be higher than it results from particles density function. This "clustering" effect could be most likely explained by fact, that particles in the shower are produced in at least pairs. Thus, when originating not very high in the atmosphere they should hit the ground close to each other as they may not have time to travel further. In this work this results are included in calculations. The function which scale the probability of multiple signals is a function of particles density and it impacts signals of 2 and more coincidences.

To evaluate expected number of events caused by EAS an integration over all energies of primary particles, area around the devices and spherical angle of the sky must be performed. Obviously, assumption that cascades are circularly symmetrical reduces integral over horizontal angle to the factor of $2 \pi r$. Formula below shows how it is computed:

\begin{equation}
    \langle N\left(k\right) \rangle = \int_{0}^{r_{max}} \int_{E_{min}}^{E_{max}} \int_0^{\frac{\pi}{2}} Q\left(n, k, P\right) 2 \pi r\ j\left(E\right) T \,d\Omega \,dE \,dr
    \label{eq:N}
\end{equation}

\begin{figure}[h!]
    \centering
    \begin{subfigure}[b]{0.47 \textwidth}
        \centering
        \includegraphics[width=\textwidth]{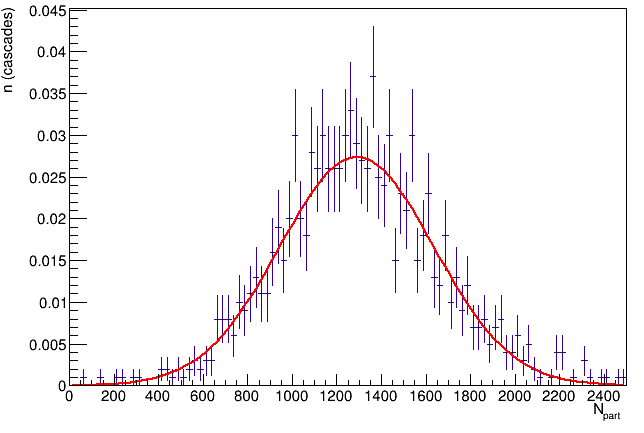}
        \caption{}
        \label{fig:N_muons}
    \end{subfigure}
    \begin{subfigure}[b]{0.47 \textwidth}
        \centering
        \includegraphics[width=\textwidth]{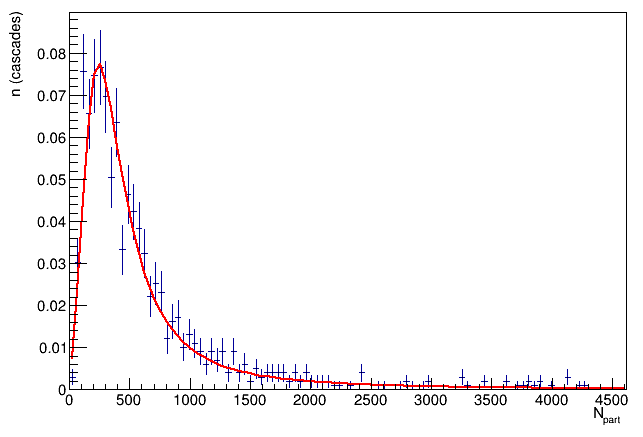}
        \caption{}
        \label{fig:N_gamma}
    \end{subfigure}
    \caption{Normalised distribution of produced particles for \ref{fig:N_muons} muons with fitted Gaussian distribution function and \ref{fig:N_gamma} photons with fitted Landau distribution function. Both plots depicts vertical showers with primary energy $E = 100\ TeV$.}
    \label{fig:N_distrib}

\end{figure}

During numerical integration the number of produced particles in cascades was chosen randomly according to distributions found earlier. Examples of those are presented in figure \ref{fig:N_distrib}.

Of course the integration must be performed in certain limits, thus the primary cosmic-rays energy range and maximal distance from the shower must be chosen. Integration up to $\frac{\pi}{2}$ over zenith angle is obvious as it was assumed that particles from all directions can be detected. Criteria of choice of the maximal distance may be different and in this work three were considered. First one was a radius in which $95 \%$ percent of particles produced by the cascade are included $R_{prc}$. Another one is a radius in which shower particles density is greater than the average background $R_{rho}$. Last one was just a fixed value $R_{max}$, chosen in a way that cuts off the region in which particles density is relatively low. First two of those quantities were analysed and characterised as a function of energy of primary particle.

\section{Comparison with simpler method}
\label{comparison with simpler method}

To check if the performed analysis was done right another approach to the problem was tested. As muons have been studied for many years, an approximate function for its density on the ground level can be found in the literature \cite{particle1998review}. It depends on total number of produced muons, which is a function of the energy of primary particle, and distance from the centre of the shower. This was also used for electromagnetic component of the cascades, however this extrapolation is very rough as electrons and photons in EAS behave very differently from muons. Thus, used formula is as follows:

\begin{equation}
    \rho\left(r, \theta, N_{part}\right) = F_{\theta}\left(\theta\right) \cdot \frac{1.25 N_{part}(E)}{2\pi \Gamma\left(1.25\right)} \left(\frac{1}{320}\right)^{1.25} r^{-0.75} \left(1 + \frac{r}{320}\right)^{-2.5}
    \label{eq:rho_approx}
\end{equation}

Where $F_{\theta}$ and further steps in the analysis are the same as described in \ref{signals from the cascades}. This comparison should help to judge if analysis described earlier, which is more general and gives more information about the behaviour of the system, is reasonable and does not consist any unrealistic assumptions.

\section{Results}
\label{results}

Finally, in this sections results of above analysis calculated for exemplary system are presented. Assumed properties of considered system are as follows. It is a system built of 4 devices, namely Cosmic Watches \cite{axani2018cosmicwatch} which surface area is $25\ cm^2$ for each one. Coincidence time is set to $200\ ns$ and efficiency of every device is $95 \%$. Particles that gives signal are muons, electrons, positrons and photons which energy is greater than $0.3\ GeV$. The choice of above parameters are not arbitrary. They are meant to be the same as in the system tested by prof. Tadeusz Wibig in his work \cite{Karbowiak_2020}, thus the time of measurement has to be taken equal to this in the performed experiment which is $7$ days. However, there are some uncertainties like which particles really gives signal as well as efficiency and $f_{bg}$, here chosen arbitrary.

Assumptions described earlier imply that the flux of background particles should be around $163\ particles\ m^{-2}\ s^{-1}$. Frequency of non cosmic background noise was chosen to be $0.1\ Hz$.

To perform final results also integration limits had to be chosen. Reasonable choice of energy limits was range between $1\ TeV$ and $10^7\ TeV$, as less energetic particles almost does not produce cascades and more energetic are very unlikely to occur in such measurement time. Maximal distance from the system in which showers were considered was $R_{prc}$ described in \ref{signals from the cascades}.

Because chronically first assumptions was that only muons should give signal also this case is presented here to show the difference in final results that yields from such changes in assumptions. After carrying all steps of the analysis the  results are as follows:

\begin{table}[h]
    \centering
    \begin{tabular}{|c||c|c||c|c||c|c||c|} \hline
        & \multicolumn{2}{|c||}{Background} & \multicolumn{2}{|c||}{Analysis} & \multicolumn{2}{|c||}{Simpler method} & Measurement \\ \cline{2-8}
        & Only $\mu$ & $\mu$, $e^{\pm}$, $\gamma$ & Only $\mu$ & $\mu$, $e^{\pm}$, $\gamma$ & Only $\mu$ & $\mu$, $e^{\pm}$, $\gamma$ & \\ \cline{2-7}
        $k$ & $\langle N (k) \rangle_{bg}$ & $\langle N (k) \rangle_{bg}$ & $\langle N (k) \rangle$ & $\langle N (k) \rangle$ & $\langle N (k) \rangle$ & $\langle N (k) \rangle$ & $N_{data} (k)$  \\ \hline \hline
        1 & $860000$ & $1.17 \cdot 10^{6}$ & $41000$ & $64000$ & $59000$ & $178000$ & - \\ \hline
        2 & $0.092$ & $0.17$ & $0.179$ & $426$ & $0.213$ & $63$ & 94 \\ \hline
        3 & $\sim 10^{-9}$ & $\sim 10^{-8}$ & $0.0182$ & $31$ & $0.003$ & $0.406$ & 2\\ \hline
        4 & $\sim 10^{-17}$ & $\sim 10^{-16}$ & $0.0068$ & $21$ & $0.0006$ & $0.143$ & 1 \\ \hline

    \end{tabular}
    \caption{Results of the calculations for cascades and background signals compared with measurements \cite{Karbowiak_2020}.}
    \label{tab:results}
\end{table}

As presented in table \ref{tab:results}, measured number of events lay somewhere between results of calculations which considered only muons and those that included also electromagnetic component. Additional information about the measurement is that it was held on the windowsill, which when assuming that electromagnetic component of EAS can pass only through the window could be taken into account in calculations. Considering that only $30 \%$ of cascades for which $\theta \geq 15^o$ contribute $e^{\pm}$ and $\gamma$ to density, the results are as follows:
   
\begin{table}[h]
    \centering
    \begin{tabular}{|c||c|c|c||c|} \hline
        $k$ & $\langle N (k) \rangle_{\mu}$ & $\langle N (k) \rangle_{\mu, e, \gamma}$ &  $\langle N (k) \rangle_{window}$ & $N_{data} (k)$ \\ \hline \hline
        2 & $0.179$ & $426$ & $99.7$ & $94$ \\ \hline
        3 & $0.0182$ & $31$ & $5.9$ & $2$ \\ \hline
        4 & $0.0068$ & $21$ & $3.9$ & $1$ \\ \hline

    \end{tabular}
    \caption{Results of the calculations for cascades and background signals compared with measurements \cite{Karbowiak_2020}}.
    \label{tab:FinalResults}
\end{table}

\section{Summary}
\label{summary}

There are several important conclusions that can be made from presented results. As expected the average number of coincidence signals caused by EAS is significantly higher than for the background. Thus, the level of confidence that certain event indicates occurrence of the cascade in a close surrounding of the system should be high. Assuming that only muons give signal is clearly wrong, however not all electrons and photons should be able do it as well. It must be clarified at which energy particles of each type give signal. Possibly the electromagnetic component of the showers is blocked in the structure of the building but more measurements in different conditions have to be performed. There are still many significant assumptions and simplifications like treating all showers as originating from protons while heavier nuclei should contribute differently to the results. Background and effects like particles production in the upper layers of the building are also poorly understood and require more study. Therefore, one should treat presented results with caution.

According to this analysis, this method of detection could even provide some clues about primary energy of detected shower, which is easy to see when the expected number of events for different number of coincidence signals is plotted over energy spectrum:

\begin{figure}[h!]
    \centering
    \begin{subfigure}[b]{0.47 \textwidth}
        \centering
        \includegraphics[width=\textwidth]{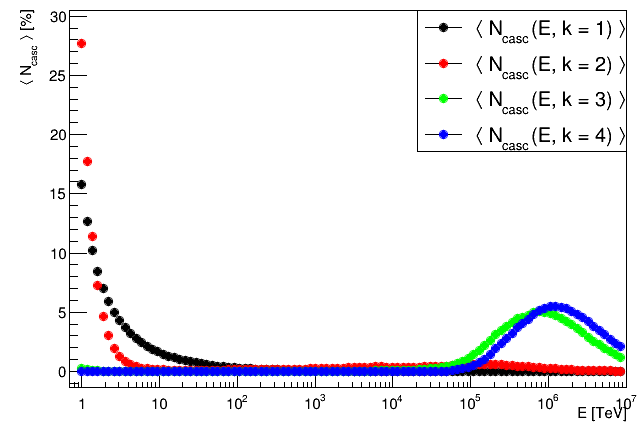}
        \caption{}
        \label{fig:NEApprox}
    \end{subfigure}
    \begin{subfigure}[b]{0.47 \textwidth}
        \centering
        \includegraphics[width=\textwidth]{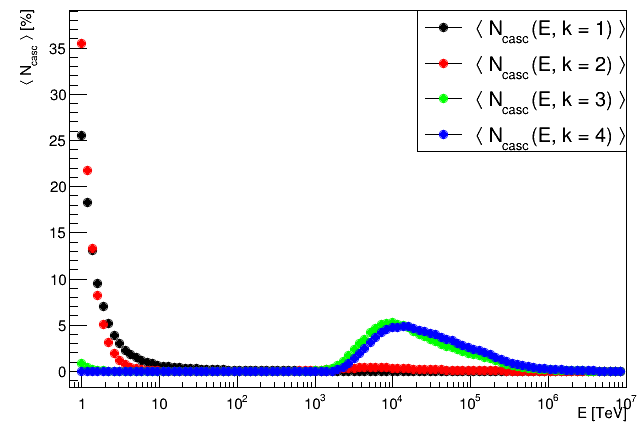}
        \caption{}
        \label{fig:NE}
    \end{subfigure}
    \caption{Expected number of coincidence signals for different energies for \ref{fig:NEApprox} simpler method and \ref{fig:NE} analysis.}
\end{figure}

Future development of this work will include more groups of heavier primary nuclei and different energy thresholds of individual components of the showers hopefully compared with more measurements.

\bibliographystyle{JHEP}
\bibliography{ICRCbibliography}

%Full authors list (ONLY FOR COLLABORATIONS)
\clearpage
\section*{Full Authors List: \Coll\ Collaboration}
%
%\noindent \textbf{Note comment afterwards:} Collaborations have the possibility to provide an authors list in xml format which will be used while generating the DOI entries making the full authors list searchable in databases like Inspire HEP. For instructions please go to icrc2021.desy.de/proceedings or contact us under icrc2021proc@desy.de.\\
%
%Authors

\scriptsize
\noindent
Jerzy Pryga$^1$,
Weronika Stanek$^2$,
Piotr Homola$^3$,
Tadeusz Wibig$^{19}$,
Oleksandr Sushchov$^3$, 
Peter Kovacs$^8$,
Vahab Nazari$^{17,3}$,
Jilberto Zamora-Saa$^{20}$,
Dmitriy Beznosko$^3$,
Mikhail~V.~Medvedev$^{10,11}$,
Jaros\l{}aw Stasielak$^{3}$,
Bartosz {\L}ozowski$^9$,
Nikolai Budnev$^4$,
Alok C. Gupta$^5$,
Arman Tursunov$^{18}$,
Karel Smolek$^{16}$,
Bohdan Hnatyk$^6$,
David E. Alvarez Castillo$^{3,17}$, 
Katarzyna Smelcerz$^{12}$,
Alona Mozgova$^6$,
Micha\l{} Nied{\'z}wiecki$^{12}$,
Marcin Kasztelan$^7$,
Mat{\' i}as Rosas$^{15}$,
Krzysztof Rzecki$^{2}$,
S\l{}awomir Stuglik$^{3}$.\\

%Dariusz G{\'o}ra$^1$,
%Justyna Miszczyk$^1$,
%Maciej Pawlik$^{13,2}$,
%Manana Svanidze$^{17}$,
%Yuri Verbetsky$^{17}$, \\
\noindent
$^1$Jagiellonian University, Gołębia 24, 31-007 Kraków.\\
$^2$AGH University of Science and Technology, Mickiewicz Ave., 30-059 Kraków, Poland.\\
$^3$Institute of Nuclear Physics Polish Academy of Sciences, Radzikowskiego 152, 31-342 Krak{\'o}w, Poland.\\
$^4$Irkutsk State University, Russia.\\
$^5$Aryabhatta Research Institue of Observational Sciences (ARIES), Manora Peak, Nainital 263001, India.\\
$^6$Astronomical Observatory of Taras Shevchenko National University of Kyiv, 04053 Kyiv, Ukraine.\\
$^7$National Centre for Nuclear Research, Andrzeja Soltana 7, 05-400 Otwock-{\'S}wierk, Poland.\\
$^8$Institute for Particle and Nuclear Physics, Wigner Research Centre for Physics, 1121 Budapest, Konkoly-Thege Mikl{\'o}s {\'u}t 29-33, Hungary.\\
$^9$Faculty of Natural Sciences, University of Silesia in Katowice, Bankowa 9, 40-007 Katowice, Poland.\\
$^{10}$Department of Physics and Astronomy, University of Kansas, Lawrence, KS 66045, USA.\\
$^{11}$Laboratory for Nuclear Science, Massachusetts Institute of Technology, Cambridge, MA 02139, USA.\\
$^{12}$Department of Computer Science, Faculty of Computer Science and Telecommunications, Cracow University of Technology, Warszawska 24, 31-155  Krak{\'o}w, Poland.\\
$^{13}$ACC Cyfronet AGH-UST, 30-950 Krak{\'o}w, Poland.\\
$^{14}$Clayton State University, Morrow, Georgia, USA.\\
$^{15}$Liceo 6 Francisco Bauz{\' a}, Montevideo, Uruguay.\\
$^{16}$Institute of Experimental and Applied Physics, Czech Technical University in Prague.\\
%$^{17}$E. Andronikashvili Institute of Physics under Tbilisi State University, Georgia.\\
$^{17}$Joint Institute for Nuclear Research, Dubna, 141980 Russia.\\ 
$^{18}$Research Centre for Theoretical Physics and Astrophysics, Institute of Physics, Silesian University in Opava, Bezru{\v c}ovo n{\'a}m. 13, CZ-74601 Opava, Czech Republic.\\
$^{19}$University of {\L}{\'o}d{\'z}, Faculty of Physics and Applied Informatics, 90-236 {\L}{\'o}d{\'z}, Pomorska 149/153, Poland.\\
$^{20}$ Universidad Andres Bello, Departamento de Ciencias Fisicas, Facultad de Ciencias Exactas, Avenida Republica 498, Santiago, Chile.\\

\end{document}